\documentclass{ws-ijmpb}

\begin{document}

\markboth{Z.-Z. Zhang and L.-L. Rong} {Growing Small-World Networks
Generated by Attaching to Edges}

%
\catchline{}{}{}{}{}
%

\title{GROWING SMALL-WORLD NETWORKS GENERATED BY ATTACHING TO EDGES }

\author{ZHONG-ZHI ZHANG\footnote{Corresponding author.} \quad and \quad LI-LI RONG}

\address{Institute of System Engineering, Dalian University of
Technology,\\
2 Ling Gong Rd.,Dalian£¬116024,Liaoning,China\\
dlutzzz063@yahoo.com.cn (Z.-Z. Zhang)\\
llrong@dlut.edu.cn (L.-L. Rong)}



\maketitle

\begin{history}
\received{Day Month Year}
\revised{Day Month Year}
\end{history}

\begin{abstract}
We introduce a minimal model of growing small-world network
generated by attaching to edges. The produced network is a plane
graph which exists in real-life world. We obtain the analytic
results of degree distribution decaying exponentially with degree
and average clustering coefficient $C=\frac{3}{2}ln3-1\approx
0.6479$, which are in good agreement with the numerical simulations.
We also prove that the increasing tendency of average path length of
the considered network is a little slower than the logarithm of the
network order $N$.
\end{abstract}

\keywords{Complex networks; Small-world Networks; Disordered
systems.}

\section{Introduction}    

Many real-world systems take the form of networks---verieces
connected together by edges.\cite{AlBa02,DoMe02,Ne03,St01,Ne00}
Commonly cited examples include technological networks such as the
Internet,\cite{FaFaFa99} information networks such as the World Wide
Web,\cite{AlJeBa99} social networks such as co-author
networks\cite{Ne01} and sexual networks,\cite{LiEdAmStAb01}
biological networks such as metabolic networks\cite{JeToAlOlBa00}
and protein networks in the cell.\cite{JeMaBaOl01} Lots of empirical
researches show that real-world networks have a small-world effect:
they have both a small average path length (APL) like random
graphs\cite{ErRe60} and a large clustering coefficient. Moreover,
the clustering coefficient seems to be irrelevant to the
network size.\\
\indent How to model real-life networks with small-world properties?
In the last few years there has been a substantial amount of
interest in network structure and function within the physics
community.\cite{AlBa02,DoMe02,Ne03,St01,Ne00} The first successful
attempt to generate networks with high clustering coefficients and
small APL is that of Watts and Strogatz (WS model).\cite{WaSt98} The
WS model starts with a ring lattice with $N$ vertices in which every
vertex is connected to its first $2m$ neighbors ($m$ on either
side). The small-world model is then created by randomly rewiring
each edge of the lattice with probability $p$ such that
self-connections and duplicate edges are excluded. The rewiring
edges are called long-range edges which connect vertices that
otherwise would be part of different neighborhoods. The pioneering
article of Watts and Strogatz started an avalanche of research on
the properties of small-world networks and the WS model. A
much-studied variant of the WS model was proposed by Newman and
Watts,\cite{NeWa99a,NeWa99b} in which edges are added between
randomly chosen pairs of sites, but no edges are removed from the
regular lattice. In 1999, Kasturirangan proposed an alternative
model to WS small-world network.\cite{Ka99} The model starts also
with one ring lattice, then we add a number of extra vertices in the
middle of the lattice which are connected to a large number of sites
chosen randomly on the main lattice. This model is similar to the WS
model in that the addition of the extra vertices effectively
introduces shortcuts between randomly chosen positions on the
lattice. In fact, even in the case where only one extra vertex is
added, the model shows the small-world effect if that vertex is
sufficiently highly connected. This case has been solved exactly by
Dorogovtsev and Mendes.\cite{DoMe00} To investigate the small-world
effect further, Kleinberg has presented a generalization of the WS
model which is based on a two-dimensional square lattice.\cite{Kl00}
Recently, in order to study other mechanisms for forming small-world
networks, Blanchard and Krueger have studied a model where only
local edge formation processes are involved and no long-range random
edges at all are formed.\cite{BlKr05} Moreover, small-world networks
can be also created by various deterministic techniques:
modification of some regular graphs,\cite{CoOzPe00} addition and
product of graphs,\cite{CoSa02} and other
methods.\cite{ZhWaHuCh04,ZhRo05} Up to now, it is still significant
how to generate networks with small-world effect.
\\
\indent In this paper, we present a small-world network model using
a very simple method by attaching to edges, which was used by
Dorogovtsev \emph{et al} to generate scale-free
networks\cite{DoMeSa01,DoGoMe02} and by Zhang \emph{et al} to create
deterministic small-world networks\cite{ZhRo05}. Our brief model
displays small-world effect. We analyze the geometric
characteristics of the model both analytically and by simulations.
The network under consideration belongs to plane graphs existing in
many real-life systems.

\section{The Growing Network Model and Its Properties}

The construction of the growing small-world network is shown in Fig.
1. Now we introduce its generation algorithm. Initially $(t=2)$, the
network is a triangle consist of three vertices, s=0,1,2, each with
degree 2. At each subsequent time step, a new vertex is added, which
is attached via two links to both ends of one randomly chosen edge
that has never been selected before. The growth process is repeated
until the network grows to the desired size of $N$ vertices. We can
see easily at time $t$, the network consists of $t+1$ vertices and
$2t-1$ edges. The total degree equals $4t-2$. Thus when $t$ is large
the average vertex degree at time $t$ is equal approximately to a
constant value 4, which shows our network is sparse like many
real-life networks. It should be noted that our model is reduced to
the scale-freee network model proposed by Dorogovtsev \emph{et
al},\cite{DoMeSa01} if we have no restrictions on the edges
previously selected.

\begin{figure}[th]
\centerline{\psfig{file=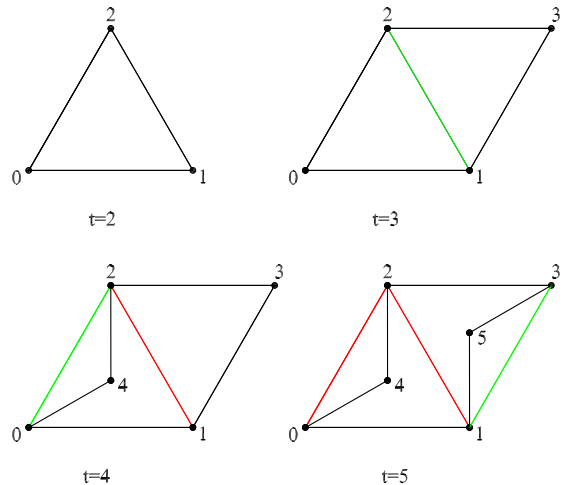,width=8cm}} \vspace*{5pt}
\caption{Scheme of the growing small-world network, showing the
first four time steps of the evolving process.}
\end{figure}

\subsection{Degree distribution}

The degree distribution is one of the most important statistical
characteristics of a network. In order to conveniently describe the
computation of network characteristics, we label vertices by their
birth times, $s=0, 1, 2,\ldots, t$, and use $p(k, s, t)$ to denote
the probability that at time $t$ a vertex created at time $s$ has a
degree $k$. Additionally, we call the edges outer edges if they have
never been chosen before. When a new vertex enter the network, it
has two outer edges. At each time step, when one outer edge is
selected, the number of outer edges of the two vertices connected by
the selected edge will decrease by one, respectively; meanwhile the
new vertex will link to them, so the number of their outer edges
will remain unchanged. Thus at time $t$, there are $t+1$ outer edges
in the network and each vertex has two, respectively. The master
equation\cite{DoMe02,DoMeSa00} governing the evolution of the degree
distribution of individual vertices has the form
\begin{equation}
p(k,s,t+1)=\frac{2}{t+1}p(k-1,s,t)+(1-\frac{2}{t+1})p(k,s,t)
\end{equation}
with the initial condition, $p(k,s={0,1,2},t=2)=\delta_{k,2}$ and
the boundary one $p(k,t,t)=\delta_{k,2}$. This describes two
possibilities for a vertex: first, with probability $\frac{2}{t+1}$,
it may get an extra edge from the new vertex and increase its own
degree by 1; and second, with the complimentary probability
$1-\frac{2}{t+1}$, the degree of the vertex may stays the same.
Eq. (1) and all the following ones are exact for all $t\geq 2$.\\
\indent The degree distribution of the network can be obtained as
\begin{equation}
P(k,t)=\frac{1}{t+1}\sum_{s=0}^{t}p(k,s,t)
\end{equation}
Using this and applying $\sum_{s=0}^{t}$ to both sides of Eq. (1),
we get the following master equation for the degree distribution,
\begin{equation}
(t+2)P(k,t+1)-(t+1)P(k,t)=2P(k-1,t)-2P(k,t)+\delta_{k,2}
\end{equation}
The corresponding stationary equation, i.e., at $t\rightarrow
\infty$, takes the form
\begin{equation}
3P(k)-2P(k-1)=\delta_{k,2}
\end{equation}
Eq. (4) implies that $P(k)$ is the solution of the recursive
equation

\begin{equation}
P(k)=\left\{\begin{array}{ll}\frac{2}{3}P(k-1)\quad for \quad k>2
\\
\frac{1}{3} \quad \quad \quad \quad \quad for \quad k=2 \\
0 \quad \quad \quad \quad \quad  otherwise\end{array}\right.
\end{equation}
giving
\begin{eqnarray}
P(k)=\frac{3}{4}\left(\frac{2}{3}\right)^{k}   \space   (k\geq 2)
\end{eqnarray}
Obviously, the degree distribution $P(k)$ is an exponential of a
power of degree $k$ (see Fig. 2), so our network can be called an
exponential one. Note that most small-world
networks including WS network\cite{BaWe00} are exponential networks.\\

\begin{figure}[th]
\centerline{\psfig{file=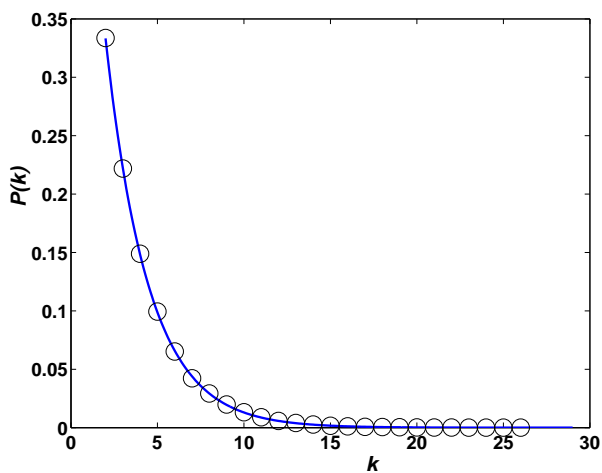,width=8cm}} \vspace*{5pt}
\caption{Degree distribution of the network with order $N=10^{5}$.
The open black circles represent the simulation results according to
our model algorithm and the solid line is the analytic calculation
value given by Eq. (6).}
\end{figure}

\subsection{Clustering coefficient}

By definition, clustering coefficient $C_{i}$ of a vertex $i$ is the
ratio of the total number $e_{i}$ of existing edges between all
$k_{i}$ its nearest neighbors and the number $k_{i}(k_{i}-1)/2$ of
all possible edges between them, ie
$C_{i}=2e_{i}/[k_{i}(k_{i}-1)]$.\cite{WaSt98} The clustering
coefficient $C$ of the whole network is the average of all
individual $C_{i}^{'}s$. In our case, we can calculate the average
clustering of the network exactly.\\
\indent When a new vertex $i$ joins the network, its degree $k_{i}$
and $e_{i}$ is 2 and 1, respectively. Each subsequent addition of an
edge to that vertex increases both $e_{i}$ and $k_{i}$ by one. Thus,
$e_{i}$ equals to $k_{i}-1$ for all vertices. So one can see that,
in this network, there is a one-to-one correspondence between the
clustering coefficient of a vertex and its degree: $C=2/k$. This
expression indicates that the local clustering scales as $C(k)\sim
k^{-1}$, which is similar to that observed in some other models.
\cite{ZhRo05,DoMeSa01,DoGoMe02} Thus, the clustering coefficient $C$
of the whole network is given by
\begin{eqnarray}
C=2\sum_{k=2}^{\infty}\frac{1}{k}P(k)=\frac{3}{2}\sum_{k=2}^{\infty}\frac{1}{k}\left(\frac{2}{3}\right)^{k}
=\frac{3}{2}ln3-1\approx 0.6479
\end{eqnarray}
So in the limit of large $t$ the clustering coefficient is large
(Fig. 3). It is worth noting that this value is higher than for
regular lattices with the same order and
average vertex degree.\cite{Ne00} \\

\begin{figure}[th]
\centerline{\psfig{file=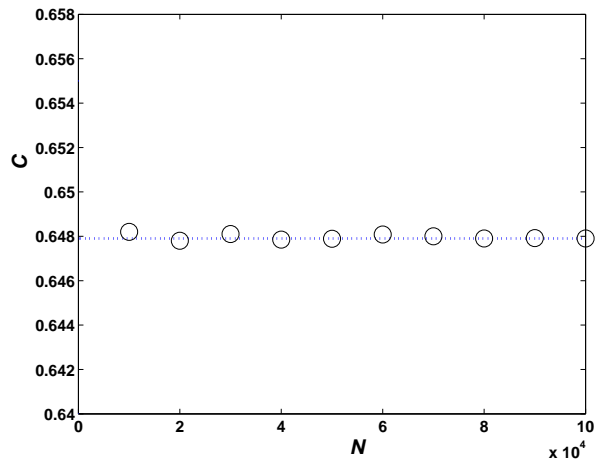,width=8cm}} \vspace*{5pt}
\caption{Clustering coefficient $C$ versus network order $N$. The
open black circles denote the simulated results, while the dotted
line is the predicted value given by Eq. (7).}
\end{figure}

\subsection{Average path length}

The average path length of a network  is defined as the number of
edges in the shortest path between two vertices averaged over all
pairs of vertices. Below, we will discuss the APL of this growing
network using the approach similar to that presented by Zhou
\emph{et al}.\cite{ZhYaWa04}

First we note that it is not very difficult to prove that for this
network and for any two arbitrary vertices $i$ and $j$ each shortest
path from $i$ to $j$ does not pass through any vertices $k$
satisfying that $k> \max(i, j)$.

  If $d(i,j)$ denotes the distance between $i$ and $j$, we introduce
the the total distance of our network with order $N$ as $\sigma(N)$
\begin{equation}
\sigma(N) = \sum_{0\leq i< j \leq N-1}d(i,j)
\end{equation}
and we denote the APL by $L(N)$, defined as:
\begin{equation}
L(N) ={2\sigma(N)\over N(N-1)}
\end{equation}
According to the former remark, the addition of new vertices will
not affect the distance between those already existing, so we have:
\begin{equation}
\sigma(N+1) = \sigma(N)+ \sum_{i=0}^{N-1}d(i,N)
\end{equation}
Assume that the vertex $N$ is added joining the edge $\mathbb{E}$
composed of vertices $w_1,w_2$, then Eq. (10) can be rewritten as:
\begin{equation}
\sigma(N+1) = \sigma(N)+ \sum_{i=0}^{N-1}(D(i,w)+1)=\sigma(N)+N+
\sum_{i=0}^{N-1}D(i,w)
\end{equation}
where $D(i,w)=\min\{d(i,w_1),d(i,w_2)\}$. Constricting the edge
$\mathbb{E}$ continuously into a single vertex $w$ (here we assume
that $w\equiv w_1$), we have $D(i,w)=d(i,w)$. Since
$d(w_1,w)=d(w_2,w)=0$, Eq. (11) can be rewritten as:
\begin{equation}
\sigma(N+1) =\sigma(N)+N+ \sum_{i\in \Gamma}d(i,w)
\end{equation}
where $\Gamma= \{0,1, 2,\cdots ,N-1\}-\{w_1,w_2\}$ is a vertex set
with cardinality $N-2$. The sum $\sum_{i\in \Gamma}d(i,w)$ can be
considered as the total distance from one vertex $w$ to all the
other vertices in our model with order $N-1$, which can be roughly
approximated in terms of $L(N-1)$:
\begin{equation}
\sum_{i\in \Gamma}d(i,w)\approx (N-2) L(N-1)
\end{equation}
Note that, as $L(N)$ increases monotonously with $N$, it is clear
that:
\begin{equation}
(N-2)L(N-1) = {2\sigma(N-1) \over N-1} < {2\sigma(N) \over N}
\end{equation}
Combining (11), (12) and (13), one can obtain the inequation:
\begin{equation}
\sigma(N+1) < \sigma(N) +N + {2\sigma(N) \over N}
\end{equation}
From (15), the variation of $\sigma(N)$ would be given by
\begin{equation}
{d\sigma(N) \over dN} = N + {2\sigma(N) \over N}
\end{equation}
This equation leads to
\begin{equation}
\sigma(N) = N^2\ln N + \alpha
\end{equation}
where $\alpha$ is a constant. As $\sigma(N) \sim N^2\ln N $, we have
$L(N) \sim \ln N$. Note that as we have deduced Eq. (17) from an
inequality, then $L(N)$ increases at most as $\ln N$ with $N$ (see
Fig. 4). Therefore, there is a desired slow growth of APL with
network size $N$.\\
\indent Based on the discussions above, the discussed network is a
sparse one with high clustering and low APL, which is obvious a
small-world network with the properties similar to other small-world networks.\cite{WaSt98,NeWa99a,NeWa99b,Ka99,DoMe00,Kl00,CoOzPe00,CoSa02,ZhRo05}\\

\begin{figure}[th]
\centerline{\psfig{file=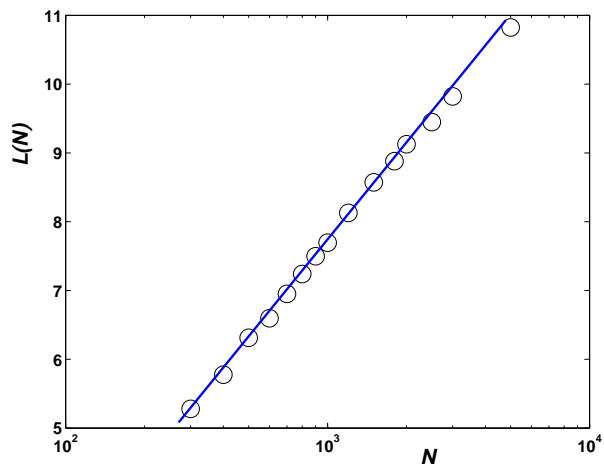,width=8cm}} \vspace*{5pt}
\caption{Average path length $L(N)$ versus network order $N$. The
black circle dots denote the simulated results. It is obvious that
$L(N)$ grows logarithmically with $N$ or more slowly, which agrees
well with the analytic results.}
\end{figure}

\section{Conclusion and Discussion}

In summary, we have proposed a simple model of growing small-world
networks. We obtain analytically some properties of the network,
which are in good agreement with the simulation results. In spite of
the simplicity of the considered model, the results are close to
those for usual small-world
networks.\cite{WaSt98,NeWa99a,NeWa99b,Ka99,DoMe00,Kl00,CoOzPe00,CoSa02,ZhRo05}
It should be mentioned that the model under consideration is a plane
graph which can be drawn on a plane without edges crossing and
attracts little attention of physicists.\cite{ZhYaWa04} Many
real-life networks are plane graphs for technical or natural
requirements, such as layout of printed circuits and vein networks
clinging to cutis. We hope our model would help to understand some
properties of real-world plane networks and be a catalyst for
further studies.

\section*{Acknowledgements}

This research was supported by the National Natural Science
Foundation of China under Grant No. 70431001.


\begin{thebibliography}{0}



\bibitem{AlBa02}
R. Albert and A.-L. Barab\'asi, {\it Rev. Mod. Phys.} {\bf 74}, 47
(2002).



\bibitem{DoMe02}
S. N. Dorogvtsev and J. F. F. Mendes, {\it Adv. Phys.} {\bf 51},
1079 (2002).


\bibitem{Ne03}
M. E. J. Newman, {\it SIAM Review} {\bf 45}, 167 (2003).

\bibitem{St01}
S. H. Strogatz, {\it Nature} {\bf 410}, 268 (2001).

\bibitem{Ne00}
M. E. J. Newman,  {\it J. Stat. Phys.}  {\bf 101}, 819 (2000).

\bibitem{FaFaFa99}
M. Faloutsos, P. Faloutsos and C. Faloutsos, {\it Comput. Commun.
Rev.} {\bf 29}, 251 (1999).


\bibitem{AlJeBa99}
R. Albert, H. Jeong and A.-L. Barab\'asi, {\it Nature} {\bf401}, 130
(1999).


\bibitem{Ne01}
M. E. J. Newman, {\it Proc. Natl. Acad. Sci. U.S.A.} {\bf 98}, 404
(2001).



\bibitem{LiEdAmStAb01}
F. Liljeros, C. R. Edling, L. A. N. Amaral, H. E. Stanley and Y.
\AA{berg} {\it Nature} {\bf 411}, 907 (2001).


\bibitem{JeToAlOlBa00}
H. Jeong, B. Tombor, R. Albert, Z. N. Oltvai and A.-L. Barab\'asi,
{\it Nature} {\bf407}, 651 (2000).

\bibitem{JeMaBaOl01}
H. Jeong, S. Mason, A.-L. Barab\'asi and Z. N. Oltvai, {\it Nature}
{\bf 411}, 41 (2001).

\bibitem{ErRe60}
P. Erd\"os and A. R\'enyi, {\it Publ. Math. Ins. Hung. Acad. Sci.}
{\bf 5}, 17 (1960).

\bibitem{WaSt98}
D. J. Watts and S. H. Strogatz, {\it Nature} {\bf 393}, 440 (1998).


\bibitem{NeWa99a}
M. E. J. Newman and D. J. Watts, {\it Phys. Lett.} {\bf A263}, 341
(1999).

\bibitem{NeWa99b}
M. E. J. Newman and D. J. Watts, {\it Phys. Rev.} {\bf E60}, 7332
(1999).

\bibitem{Ka99}
R. Kasturirangan, {\it Preprint} cond-mat/9904055.

\bibitem{DoMe00}
S. N. Dorogvtsev and J. F. F. Mendes, {\it Europhys. Lett.} {\bf
50}, 1 (2000).

\bibitem{Kl00}
J. Kleinberg, {\it Nature} {\bf 406}, 845 (2000).

\bibitem{BlKr05}
P. Blanchard and T. Krueger, {\it Phys. Rev.} {\bf E 69}, 046136
(2005).

\bibitem{CoOzPe00}
F. Comellas, J. Oz\'on, and J.G. Peters, {\it Inf. Process. Lett.},
{\bf 76}, 83 (2000).

\bibitem{CoSa02}
F. Comellas and M. Sampels, {\it Physica} {\bf A309}, 231 (2002).

\bibitem{ZhWaHuCh04}
T. Zhou, B. H. Wang, P. M. Hui and K. P. Chan, {\it Preprint}
cond-mat/0405258.

\bibitem{ZhRo05}
Z. Z. Zhang and L. L. Rong, {\it Preprint} cond-mat/0502335.

\bibitem{DoMeSa01}
S. N. Dorogovtsev, J. F. F. Mendes and A. N. Samukhin, {\it Phys.
Rev.} {\bf E63}, 062101 (2001).

\bibitem{DoGoMe02}
S. N. Dorogovtsev, A. V. Goltsev and J. F. F. Mendesm, {\it Phys.
Rev.} {\bf E65}, 066122 (2002).

\bibitem{DoMeSa00}
S. N. Dorogovtsev, J. F. F. Mendes and A. N. Samukhin, {\it Phys.
Rev. Lett.} {\bf 85}, 4633 (2000).

\bibitem{BaWe00}
A. Barrat, and M. Weigt, {\it Eur. Phys. J.} {\bf B13}, 547 (2000).

\bibitem{ZhYaWa04}
T. Zhou, G. Yan and B. H. Wang, {\it Phys. Rev.} {\bf E71}, 046141
(2005).


\end{thebibliography}
\end{document}